\documentclass[conference]{IEEEtran}
\IEEEoverridecommandlockouts
\usepackage{placeins}
\usepackage{amsmath,amssymb,amsfonts}
\usepackage{algorithmic}
\usepackage{graphicx}
\usepackage{textcomp}
\usepackage{xcolor}
\usepackage{microtype}
\usepackage{hyperref}
\hypersetup{hidelinks} 
\usepackage{epsfig}
\usepackage{algorithm}
\usepackage{mdwlist, multicol}
\usepackage{array, booktabs, enumitem, multirow}
\usepackage{float}
\def\BibTeX{{\rm B\kern-.05em{\sc i\kern-.025em b}\kern-.08em
		T\kern-.1667em\lower.7ex\hbox{E}\kern-.125emX}}
\begin{document}
	
	\title{Leveraging Causal Inference for Explainable Automatic Program Repair}
	
	\author{\IEEEauthorblockA{Jianzong Wang, Shijing Si$^*$, Zhitao Zhu, Xiaoyang Qu, Zhenhou Hong and Jing Xiao}
		\IEEEauthorblockA{Ping An Technology (Shenzhen) Co., Ltd., Shenzhen, China\\
			Email: jzwang@188.com, shijing.si@outlook.com, andyzzt@mail.ustc.edu.cn, \\quxiaoy@gmail.com,  hongzhenhou168@pingan.com.cn, xiaojing661@pingan.com.cn}
		\IEEEauthorblockN{\thanks{\IEEEauthorrefmark{1}Corresponding author: Shijing Si, \texttt{sishijing204@pingan.com.cn}.}}
	}
	
	\maketitle
	
	\begin{abstract}
		Deep learning models have made significant progress in automatic program repair. However, the black-box nature of these methods has restricted their practical applications. To address this challenge, this paper presents an interpretable approach for program repair based on sequence-to-sequence models with causal inference and our method is called CPR, short for causal program repair. Our CPR can generate explanations in the process of decision making, which consists of groups of causally related input-output tokens. Firstly, our method infers these relations by querying the model with inputs disturbed by data augmentation. Secondly, it generates a graph over tokens from the responses and solves a partitioning problem to select the most relevant components. The experiments on four programming languages (Java, C, Python, and JavaScript) show that CPR can generate causal graphs for reasonable interpretations and boost the performance of bug fixing in automatic program repair.
	\end{abstract}
	
	\begin{IEEEkeywords}
		Automated Program Repair, Program Analysis, Sequence-to-sequence Model, Causal Inference, Interpretability
	\end{IEEEkeywords}
	
	\section{Introduction}
	\label{sec:intro}
	The goal of automatic program repair (APR) techniques is to automatically correct bug(s) in a piece of existing troublesome source code. Many researches have been devoted on applying machine learning techniques to APR ~\cite{monperrus2018automatic}. Given the similarity between a program repair task and generic natural language processing (NLP) tasks such as sequence-to-sequence (Seq2Seq) learning and machine translation \cite{sundararaman2021syntactic}, there has been a lot of work on applying machine learning for program repair \cite{tufano2018empirical,chen2019sequencer,li2020dlfix} and \cite{lutellier2020coconut} in recent years. Similar techniques have been applied to code related tasks \cite{iyer2016summarizing,ling2016latent,yin2017syntactic}.
	
	Although deep learning based Seq2Seq models have achieved overwhelming success in APR tasks, there still remains a lot of potential for improvement \cite{nguyen2019automatic,ye2021comprehensive}. One major concern is the interpretability of the deep  Seq2Seq models, which is caused by the complicated nature of model architectures \cite{dong2017improving,falissard2022neural}. The interpretability of deep models is often categorized into two types: 1.) model interpretability, which attempts to make the deep neural architecture itself interpretable and transparent, and 2.) prediction interpretability, which aims to explain particular predictions of the deep model ~\cite{lei2016rationalizing}. Once a deep APR model has achieved these two aspects of interpretability, it becomes trustworthy, transparent, and controllable, which surely makes deep models more useful for practical deployment. 
	
	To improve the interpretability of APR models, this paper introduces causal inference, which has shown great promise in discovering causal relations in deep learning \cite{alvarez2017causal}.
	Simply combining APR with causal inference can lead to a few significant challenges.
	This paper aims to address three important unsolved issues about how to use causal inference in APR models. Because transparency in the deep APR models is often very restrictive and challenging to achieve \cite{o2020automatic}, the first issue is how can we alleviate this difficult situation in real application. The second issue is how to define the causal graph in the APR task. The third issue is what kind of performance can be achieved by APR model combined with causal inference.
	
	For the first issue, in this work we concentrate on prediction interpretability
	rather than model transparency. From the machine learning community, prediction interpretability can be sought more easily with existing mehtods, like Monte Carlo Dropout, Ensemble or Bayesian deep neural networks \cite{liu2020simple}. 
	In a typical APR problem, the inputs are the original code and comments, the output is the predicted repaired code. To tackle the second issue, we can define the causal graph since we are focusing on the interpretable relationship between input and output. In a causal graph, the context is the input of code and comments, the outcome is the predicted repaired code, and the confounder is the data disturbance, i.e., the data augmentation. For the third challenge, we alter the potential confounder, which is one kind of text data augmentation, and observe the effect of APR with causal inference.
	
	Specifically, we propose to utilize data augmentation strategy to discover the causal relations between the input source (that is, the code and comments) and the corrected bugs, which can
	improve the prediction interpretability of APR models. 
	Data augmentation methods are used to disturb the original inputs, which are then fed into the deep Seq2Seq model to infer the causal relation between the input and output tokens in the similar manner as \cite{alvarez2017causal}.  \cite{alvarez2017causal} utilized a Variational auto-Encoder for data disturbation, which is computationally costly and hard to control. Our method is called CPR, short for Causal Program Repair, which leverages text data augmentation and easy to implement.
	
	\begin{figure*}[htbp]
		\begin{minipage}[b]{0.33\linewidth}
			\centering
			\includegraphics[width=0.81\linewidth]{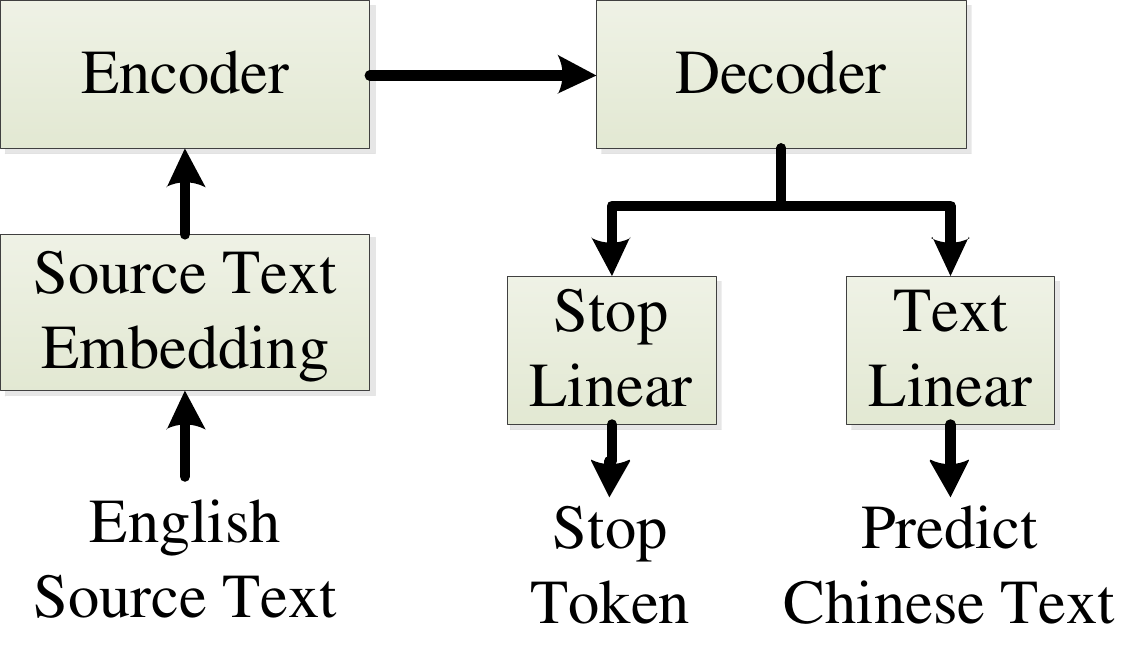}
			\\(a) The framework of Seq2Seq model in neural machine translation (NMT).
		\end{minipage}
		\begin{minipage}[b]{0.33\linewidth}
			\centering
			\includegraphics[width=0.84\linewidth]{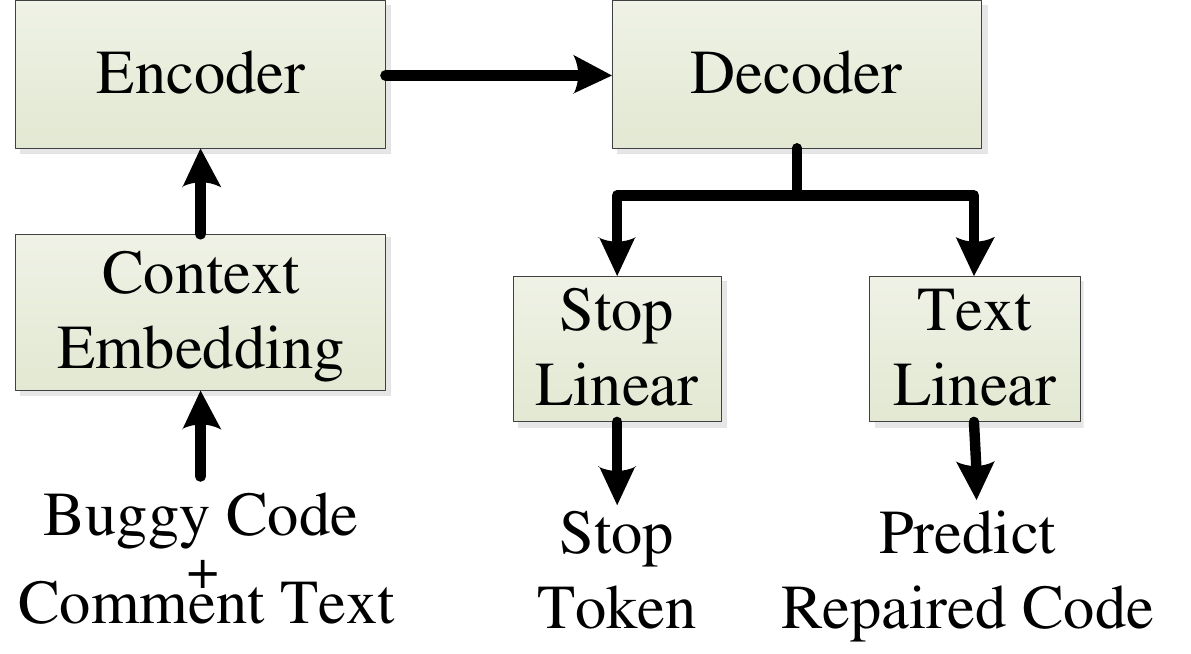}
			\\(b) The framework of Seq2Seq model in APR.
		\end{minipage}
		\begin{minipage}[b]{0.33\linewidth}
			\centering
			\includegraphics[width=\linewidth]{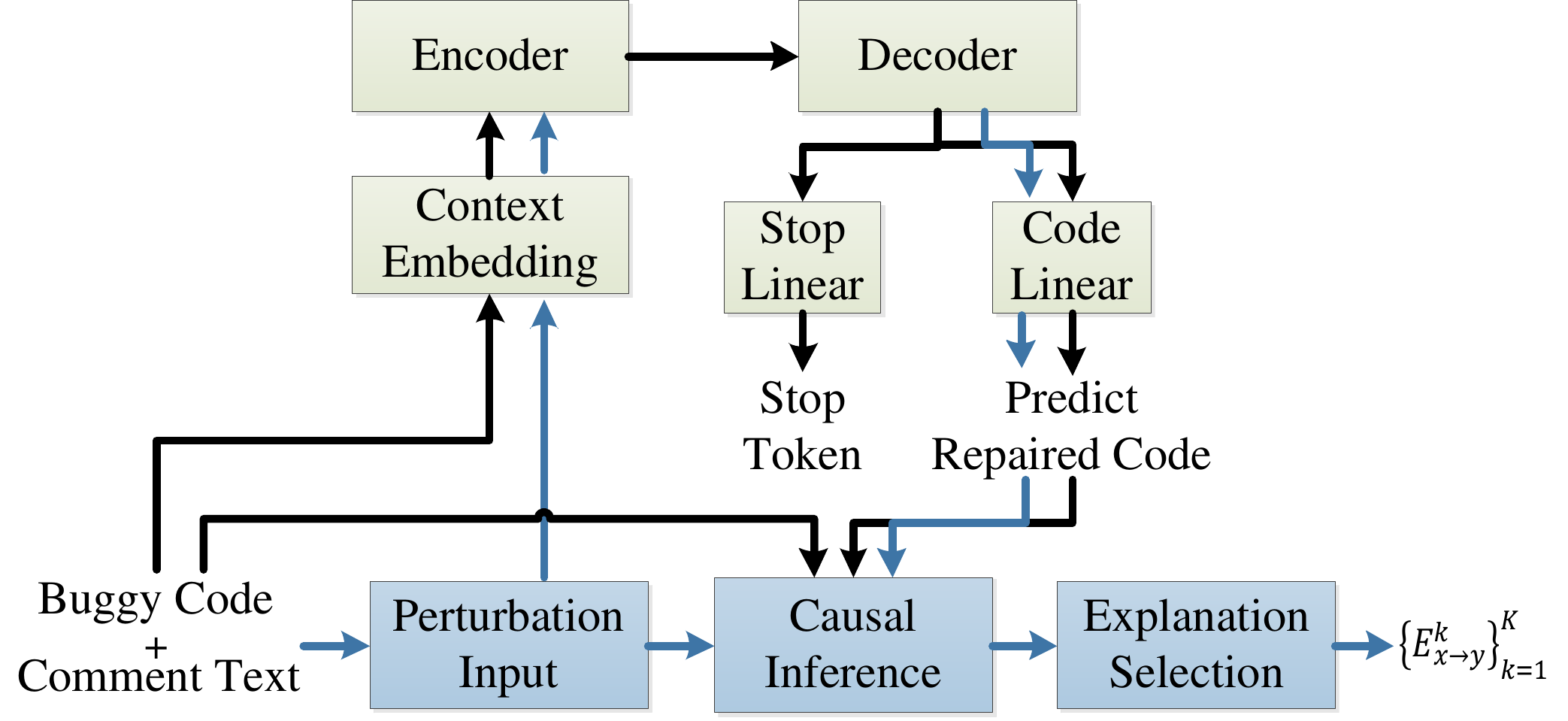}
			\\(c) The framework of Seq2Seq model in CPR.
		\end{minipage}
		\caption{Comparison of three frameworks of NMT, APR and CPR.
			The boxes and arrows in blue in (c) mean the causal parts.}
		\label{fig:nmt-coderepair-cacoderepair}
	\end{figure*}
	
	Our main contributions can be listed as follows:
	\begin{itemize}
		\item We proposed to utilize data augmentation strategy for the input perturbations required for causal analysis of program repair
		\item In the experiments, our framework can offer explanations on various Seq2Seq models in APR, which exhibits its ability to enhance the performance and provide insights into how the black-box models make predictions.
	\end{itemize}
	
	\section{Related Works}
	\label{sec:related}
	Our CPR is closely related to two lines of research: Seq2Seq model for automatic code repair and the causal inference. Therefore, we discuss the related research in the following two subsections.
	
	\subsection{Sequence-to-Sequence Models in Automatic Code Repair}
	
	There are many existing works that attempt to use machine translation techniques to automatically fix bugs. Here are the most notable methods in this area. \cite{hata2018learning} formulates the patches generation as a Seq2Seq translation problem and proposes to use a neural machine translation (NMT) model with attention-based Encoder-Decoder.
	To evaluate the performance of various APR methods, \cite{madeiral2019bears} presents BEARS, a project for collecting and storing bugs into
	an extensible bug benchmark for automatic repair studies in Java.  
	SequenceR~\cite{chen2019sequencer} leverages a Seq2Seq model by combining an encoder and a decoder architecture, where recurrent neural networks (RNNs) with LSTM gates are used for both the encoder and decoder. 
	\cite{nguyen2019automatic} propose an APR framework using formal specification and
	expression templates.
	
	Using a RNN encoder-decoder, \cite{tufano2019empirical} investigates the application of NMT for candidate patches generation.
	Applying NMT techniques and ensemble learning, CoCoNuT~\cite{lutellier2020coconut} can automatically repair programs of multiple languages in a end-to-end manner. 
	\cite{jiang2021cure} proposes a new NMT-based APR technique, called CURE, which leverages pre-training, subword tokenization and an efficient code-aware search strategy.
	\cite{campos2021automatic} proposes a real-time code fix mechanism by semantic code suggestions, which is shown to improve
	the speed of repairing faulty programs.
	More details on APR can be found in \cite{mamatha2022literature}.
	
	\subsection{Causal Inference}
	There is a very large body of work that attempts to address the issue of "explanation", but it is actually quite biased due to the different definitions of "explanation" that are of interest.
	In the field of machine learning, the area where interpretability is probably most valued is in medical applications, where the credibility of a predictive model depends heavily on its interpretability~\cite{caruana2015intelligible}. 
	With the advent of the deep learning trend, recent work has enhanced each of the two aspects of interpretability: model  transparency \cite{DBLP:conf/cvpr/MahendranV15} and model functionality \cite{DBLP:conf/ijcai/ZhaoOK20}.
	For a broad survey of interpretability in deep learning, we refer the reader to the survey \cite{doshi2017roadmap}.
	
	Model interpretability methods based on causal inference (counterfactual samples) have been increasingly applied to various scenarios. Among them, the sample-based explanation approach aims to explain the decision and judgment process of the model by finding sample examples. \cite{lei2016rationalizing,ribeiro2016should} proposes a model that justifies the prediction results using fragments of the input. One of the most typical approaches is the counterfactual explanations that interpret the model's decisions by making minimal changes to the features on the existing samples and obtaining the expected counterfactual results, and collecting these samples with minor changes. \cite{cheng2021fairfil} relies on local perturbations of the instance to explain the predictions of the black-box classifiers.
	A generic counterfactual generator with sequential control of perturbation types and positions is further proposed by \cite{DBLP:conf/acl/WuRHW20}, which can generate diverse sets of realistic counterfactuals that can be useful in various distinct applications.
	\section{Proposed Method}
	\label{sec:propsm}
	Comparing the Fig.~\ref{fig:nmt-coderepair-cacoderepair}(a), (b) and (c), we can transition from NMT to APR with causal inference, which is Fig.~\ref{fig:nmt-coderepair-cacoderepair}(c). Comparing Fig.~\ref{fig:nmt-coderepair-cacoderepair} (a) and (b), the process of APR is similar to NMT. Both input source text and target text, and the expected text is inferred through the Seq2Seq model. According to Fig.~\ref{fig:nmt-coderepair-cacoderepair}(b), we propose an APR framework with causal analysis, as illustrated by Fig.~\ref{fig:nmt-coderepair-cacoderepair}(c).
	
\begin{figure}
		\begin{minipage}[b]{0.49\linewidth}
			\centering
			\includegraphics[width=0.4\linewidth]{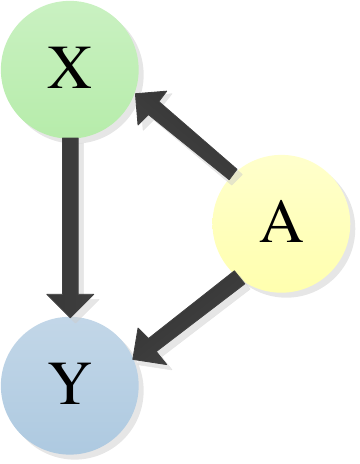}
			\\(a) The causal graph in APR task.
		\end{minipage}
		\begin{minipage}[b]{0.49\linewidth}
			\centering
			\includegraphics[width=0.9\linewidth]{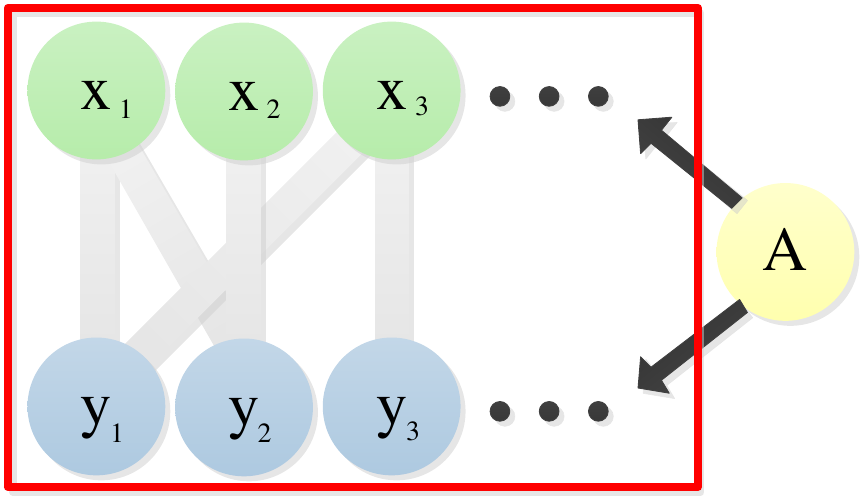}
			\\(b) The input and output connections.
		\end{minipage}
		\caption{The causal graph in APR task.}
		\label{fig:caugra}
	\end{figure}

	As demonstrated in Fig.~\ref{fig:caugra}(a), we define a causal graph. 
	In causal inference, capital X represents the treatment, capital A represents the potential confounder, and capital Y represents the outcome. In the APR problem, capital X is the input of code and comments, capital A indicates the data augmentation, and capital Y is the output of predicted repaired code. Based on the data augmentation, potential confounder A will influence X and Y, so we define the causal graph like the Fig.~\ref{fig:caugra}(a). To identify the input and output causality, we need to know the predict interpretability after we've defined the causal graph. 
	Changing the input through data augmentation methods, we may figure out how the input influences the output, as shown in Fig.~\ref{fig:caugra}(b).
	
	\begin{table*}[htbp]
		\centering
		\fontsize{9}{13}
		\selectfont
		\caption{Comparison of the fixed bugs with causal inference and without causal inference.
			``CI'' means causal inference. The effectiveness of models is measured by the number of fixed bugs
		}
		\label{tab:tablecom}
		\begin{tabular}{ccccccc}
			\toprule
			\textbf{Dataset} & \textbf{Model}
			& \textbf{Paremeters}
			& \textbf{Language}
			& \textbf{Bugs number}
			& \textbf{fixed w/o CI}
			& \textbf{fixed with CI (CPR)}\\ 
			\midrule
			Defects4J & SequenceR & 38M & Java & 393 & 12 & \textbf{20} \\
			Defects4J & FConv & 23M & Java & 393 & 12 & \textbf{21} \\
			Defects4J & CodeBERT & 675M & Java & 393 & 14 & \textbf{23} \\
			\midrule
			QuixBugs & SequenceR & 38M & Java & 40 & 7 & \textbf{13} \\
			QuixBugs & FConv & 23M & Java & 40 & 8 & \textbf{11} \\
			QuixBugs & CodeBERT & 675M & Java & 40 & 13 & \textbf{17} \\
			\midrule
			QuixBugs & SequenceR & 38M & Python & 40 & 6 & \textbf{13} \\
			QuixBugs & FConv & 23M & Python & 40 & 9 & \textbf{14} \\
			QuixBugs & CodeBERT & 675M & Python & 40 & 12 & \textbf{19} \\
			\midrule
			ManyBugs & SequenceR & 38M & C & 69 & 11 & \textbf{16} \\
			ManyBugs & FConv & 23M & C & 69 & 9 & \textbf{15} \\
			ManyBugs & CodeBERT & 675M & C & 69 & 13 & \textbf{18} \\
			\midrule
			BugAID & SequenceR & 38M & JavaScript & 12 & 4 & \textbf{6} \\
			BugAID & FConv & 23M & JavaScript & 12 & 3 & \textbf{3} \\
			BugAID & CodeBERT & 675M & JavaScript & 12 & 6 & \textbf{7} \\
			\toprule
		\end{tabular}
	\end{table*}
	
	Our approach formalizes this framework through a pipeline (sketched in Fig.~\ref{fig:nmt-coderepair-cacoderepair}(c)) that consists of three main components, each of which is described in detail in the following section: a perturbation model for locally exercising $F$, a causal inference model for inferring associations between inputs and predictions, and a selection step for partitioning and selecting the most relevant sets of associations.
	\vspace{0.3cm}
	\begin{figure}[htbp]
		\centerline{\includegraphics[width=\linewidth]{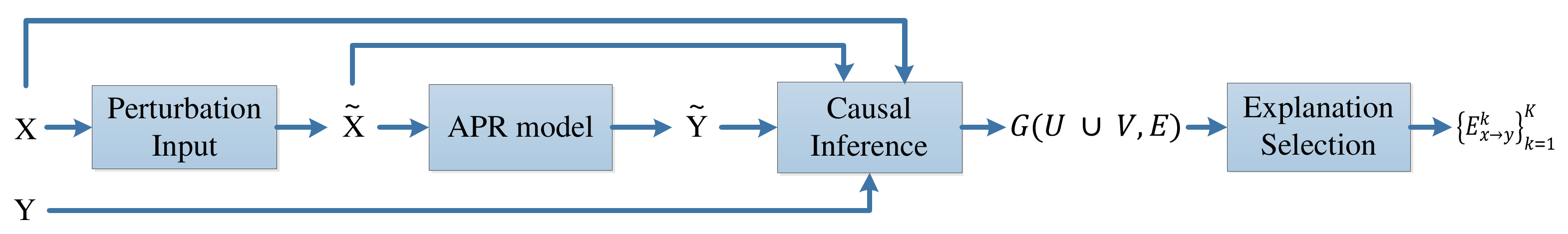}}
		\caption{The framework of CPR.}
		\label{fig:cacoframe}
	\end{figure}
	\vspace{0.5cm}

	The first step in the explainable APR approach is to obtain perturbed versions of the input: which involves changing the elements and their order in original input sentences.
	In~\cite{alvarez2017causal}, they propose a variational auto-encoder(VAE) model to perturb the text.
	However, training a VAE or bidirectional LSTM language model takes a lot of time and effort.
	
	Based on \cite{wei2019eda,edunov2018understanding}, here we use the data augmentation to generate the perturbed comment texts.
	For word perturbation:
	\begin{itemize}
		\item [1)] Synonym Replacement (SR): Select $m$ words randomly from the original sentence that does not include stop words. Then alter these words with one of closeness meaning chosen at random.
		\item [2)] Random Insertion (RI): Find a most similar meaning for a random word in the original sentence (not include stop words). Insert that most similar word into a stochastic position in the sentence. Repeat $m$ times.
		\item [3)] Random Swap (RS): Casually select two words from the original sentence and switch their positions in $m$ times.  
		\item [4)] Random Deletion (RD): With probability $p$. Delete words from the original sentence randomly. 
	\end{itemize}
	For sentence perturbation, we use Back-Translation (BT): Translating an existing example $x$ in language A into another language B and then translating it back into A to obtain an augmented example $\hat{x}$.
	
	Long sentences will perform more word perturbation since they include more words than short ones. To fair comparison, we vary the number $m$ for word perturbation based on the sentence length $l$ with the formula $m=\lfloor \alpha l \rfloor$, where $\alpha$ is a parameter that indicates the percent of the changed words in a sentence (we use $p= \alpha $ for RD). Furthermore, we generate $m_{dist}$ disturbed sentences for each original sentence. 
	
	To perturb the comment texts, the data augmentations operate the level of the words, and the symbols like \{, $:$ or $==$ e.t.c. Symbols are also an essential part of a programming language.
	
	This data augmentation strategy achieves similar outcomes but is considerably easier to employ because it does not involve the training of a language model and the usage of external datasets. We argue that it can be easily ported to other similar models.
	
	\section{Experimental Setup}
	\label{sec:exper}
	
	\subsection{Datasets}
	To train the ARP models in different programming languages, we collected corresponding training datasets for different programming languages based on the open-source datasets. For Java, we use Defects4J~\cite{just2014defects4j} and QuixBugs~\cite{lin2017quixbugs}. For Python, we use Python’s version of QuixBugs. For C, we used the ManyBugs datasets from prior work~\cite{le2015manybugs}. In order to compare with the Java, Python, and C program language, we select 12 JavaScript examples connected with ordinary bug problems in prior work (BugAID)~\cite{hanam2016discovering}.
	
	\begin{figure}[htbp]
		\begin{minipage}[b]{0.95\linewidth}
			\centering
			\includegraphics[width=1.0\linewidth]{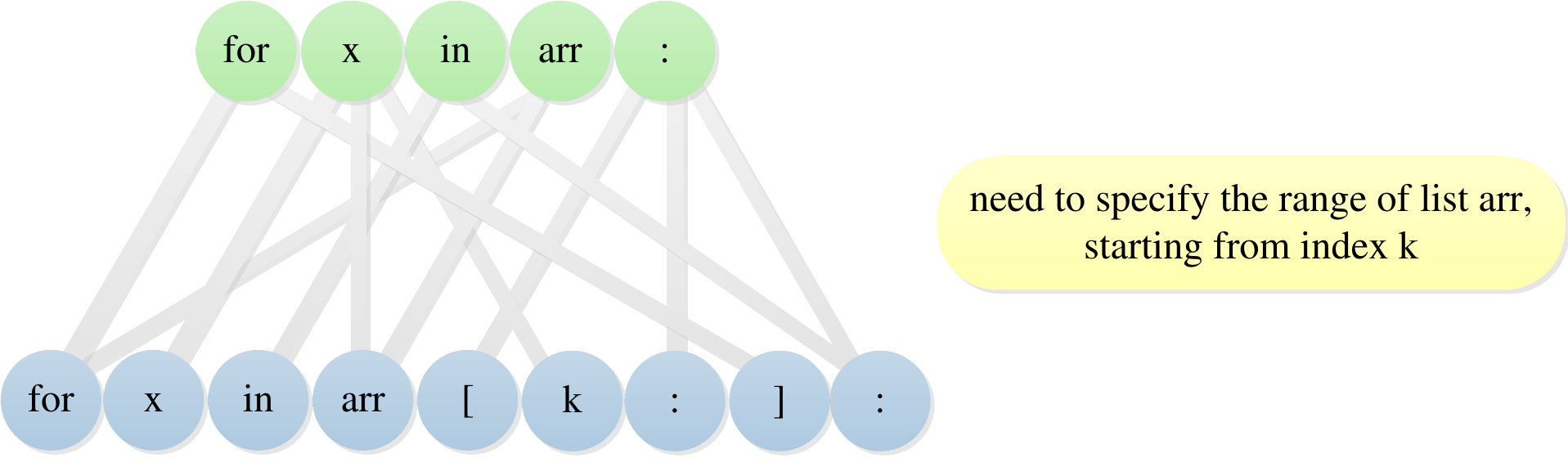}
			\\(a)
		\end{minipage}
		\begin{minipage}[b]{0.95\linewidth}
			\centering
			\includegraphics[width=1.0\linewidth]{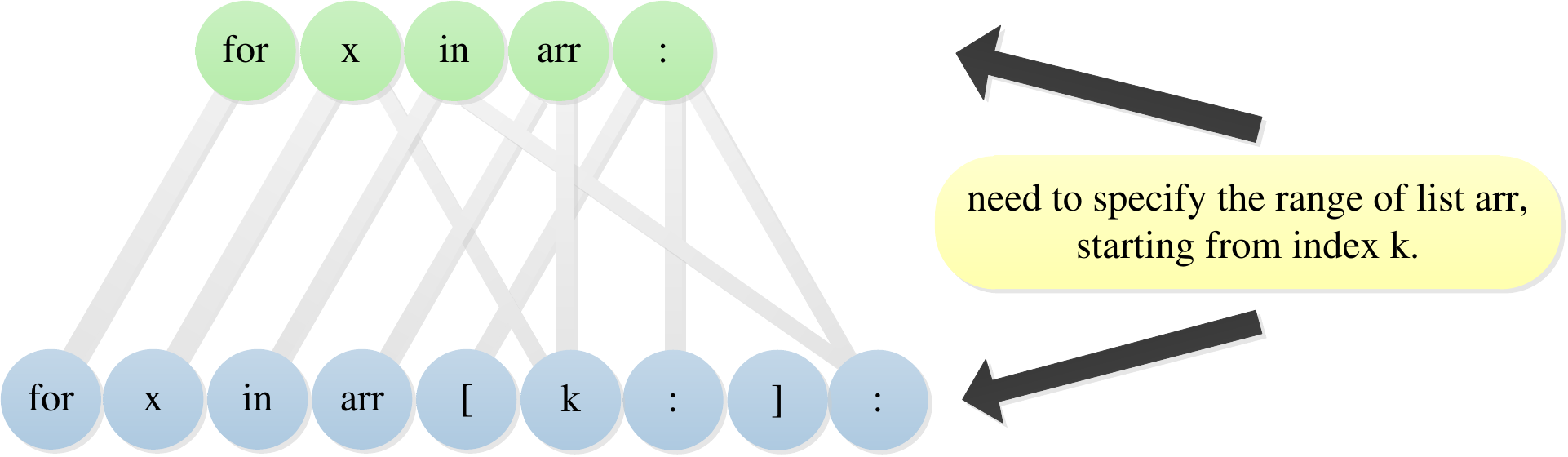}
			\\(b)
		\end{minipage}
		\caption{We inferred dependency graphs before (left) and after (right) explanation selection for the prediction.
			Our framework generates dependency estimates and explanation graphs.
			The green circles mean the buggy code, the yellow circles mean the comments text, and the blue circles mean debugged code. (a) Raw Dependencies.(b) Explanation Graph.}
		\label{fig:noex-ex}
	\end{figure}

	\subsection{Settings}
	
	\noindent\textbf{Training.}
	In order to ensure the completeness of comparison, the details of training parameters are given here. Following Fig.~\ref{fig:nmt-coderepair-cacoderepair}(c), we have two parts that need to train and inference, the APR model and the explainable graph. First, we use hyperband~\cite{li2017hyperband} to tune hyper-parameters. We restrict the range of hyper-parameters to appropriate values. The size of embedding is 64 to 512, the dimensions of convolutional layer is 32 to 256, the number of convolutional layers is 1 to 10, the size of hidden units of LSTM is 128 to 512, the size of Transformer heads is 4 to 8, the size of hidden units of Transformer is 128 to 512. The learning rate is 0.1, 0.01, 0.001, and 0.0001. Then we train APR models for one epoch on the different training sets with various hyper-parameters for tuning. We are ranking the hyper-parameters sets based on their perplexity. Perplexity is a typical metric in NLP that quantifies how well a model generates a sequence.
	We make it stop at convergence or until 20 epochs and ranking the the top-k (default k=5) models. In inference mode, we employ beam search with a beamwidth of 784.
	
	\begin{table*}[htbp]
		\centering
		\fontsize{9}{14}
		\selectfont
		\caption{Comparison of different data disturbances in our framework. ``CI'' means causal inference. The effectiveness of models is measured by the number of fixed bugs.}
		\label{tab:tableaug}
		\begin{tabular}{cccccccc}
			\toprule
			\textbf{Dataset} & \textbf{Model}
			& \textbf{Paremeters}
			& \textbf{Language}
			& \textbf{Bugs number}
			& \textbf{Data augmentation}
			& \textbf{fixed w/o CI}
			& \textbf{fixed with CI (CPR)}\\
			\midrule
			QuixBugs & SequenceR & 38M & Python & 40 & SR & 10 & \textbf{13} \\
			QuixBugs & SequenceR & 38M & Python & 40 & RI & 8 & \textbf{10} \\
			QuixBugs & SequenceR & 38M & Python & 40 & RS & 7 & \textbf{11} \\
			QuixBugs & SequenceR & 38M & Python & 40 & RD & 8 & \textbf{11} \\
			QuixBugs & SequenceR & 38M & Python & 40 & BT & 9 & \textbf{12} \\
			\midrule
			QuixBugs & FConv & 23M & Python & 40 & SR & 11 & \textbf{14} \\
			QuixBugs & FConv & 23M & Python & 40 & RI & 8 & \textbf{11} \\
			QuixBugs & FConv & 23M & Python & 40 & RS & 9 & \textbf{12} \\
			QuixBugs & FConv & 23M & Python & 40 & RD & 9 & \textbf{10} \\
			QuixBugs & FConv & 23M & Python & 40 & BT & 12 & \textbf{14} \\
			\midrule
			QuixBugs & CodeBERT & 675M & Python & 40 & SR & 14 & \textbf{18} \\
			QuixBugs & CodeBERT & 675M & Python & 40 & RI & 11 & \textbf{15} \\
			QuixBugs & CodeBERT & 675M & Python & 40 & RS & 9 & \textbf{12} \\
			QuixBugs & CodeBERT & 675M & Python & 40 & RD & 9 & \textbf{13} \\
			QuixBugs & CodeBERT & 675M & Python & 40 & BT & 13 & \textbf{19} \\
			\toprule
		\end{tabular}
	\end{table*}
	
	\noindent\textbf{Explainable graph.}
	We equally chose a data disturbance method to generate the disturbed input based on the proposed data augmentation strategy. For the explainable graph step, we use the robust clustering method of~\cite{dhillon2001co,kluger2003spectral} to generate the graph in the experiment. These bilateral clustering methods do not take uncertainty into account.
	
	\noindent\textbf{Infrastructure}
	We use the LSTM, and Transformer is running on Pytorch~\cite{paszke2019pytorch}. The implementations of FConv is provided by fairseq-py~\cite{gehring2017convolutional}. We also use the implementation of CodeBERT~\cite{feng2020codebert} running on Pytorch. Our models were trained and evaluated on an Intel Xeon E5-2695 with 4 NVIDIA V100 GPUs.
	
	\noindent\textbf{Performance}
	The average time it takes to train the APR models for one epoch during tuning is 51 minutes. Sequentially for the SequenceR, FConv, and CodeBERT, it takes 93, 67, and 253 hours to train the model until convergence. In the inference stage, producing 1000 patches for a bug brings 9 seconds on average.
	
	\subsection{Evaluation and Results}
	\label{sec:eval}
	
	\noindent\textbf{The performance with causal inference}
	Table~\ref{tab:tablecom} displays the number of bugs fixed by the different approaches with or without causal inference. For all datasets, causal inference is used to improve the effect of debugging code. We test in four languages, and both models have been improved using causal inference. On the BugAID dataset, which is for JavaScript language, We observed an intriguing result. The debugs number for all three models is relatively less, and for FConv, the causal inference is not working. For SequenceR and CdoeBERT, the effect of improvement is not obvious. We believe that the main reason is that the amount of data is too small. 
	
	Table~\ref{tab:tableaug} depicts the different consequences of different data disturbance approaches in our framework. The RI, RS, and RD approaches would degrade the performance of APR with causal inference. In the FConv model with the RD method, its performance is almost the same as that of a model without causal inference. Likewise, the CodeBERT with the RS method, which reduces the number of fixed codes to 12, achieves the same performance without causal inference. When using the APR model with causal inference, the SR and BT methods are the best choices. It demonstrates that the SR and BT methods have the best effect in the APR model with causal inference. It helps when the APR model is with the explainable graph because the SR and BT methods can maintain the original meaning of comments.
	
\begin{figure}[htbp]
		\begin{minipage}[b]{0.49\linewidth}
			\centering
			\includegraphics[width=1.09\linewidth]{fig/explaing-other-001-3-eps-converted-to.pdf}
			\\(a)
		\end{minipage}
		\begin{minipage}[b]{0.49\linewidth}
			\centering
			\includegraphics[width=0.85\linewidth]{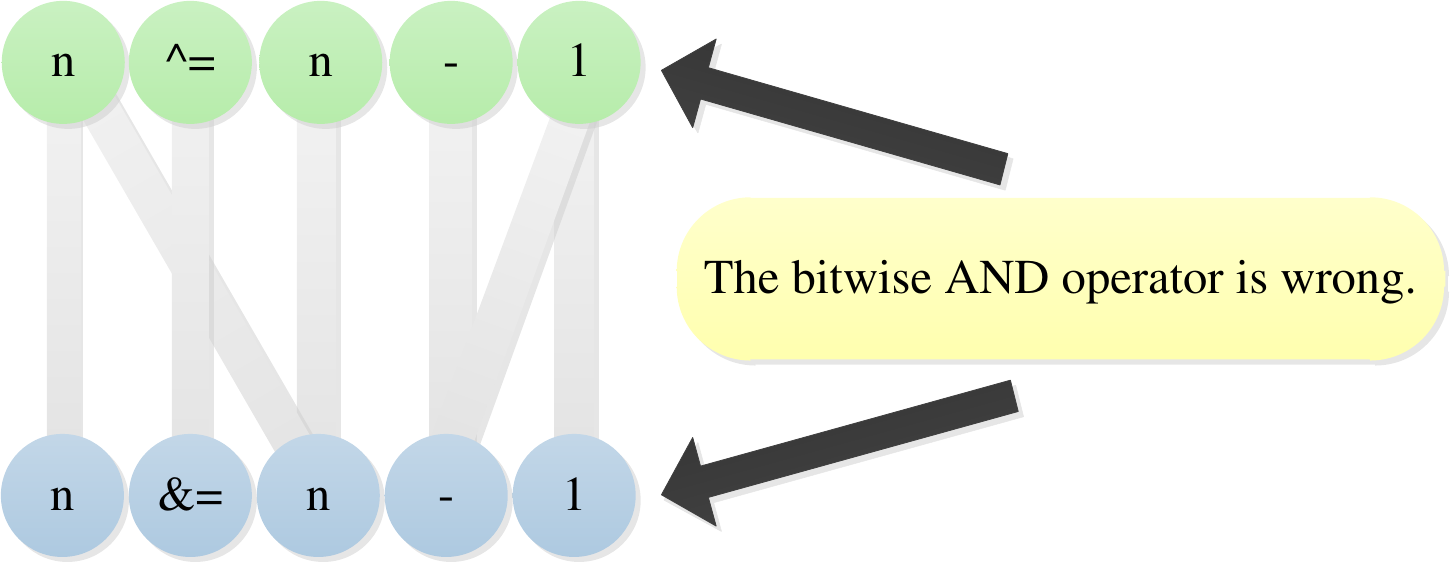}
			\\(b)
		\end{minipage}
		\begin{minipage}[b]{0.49\linewidth}
			\centering
			\includegraphics[width=0.99\linewidth]{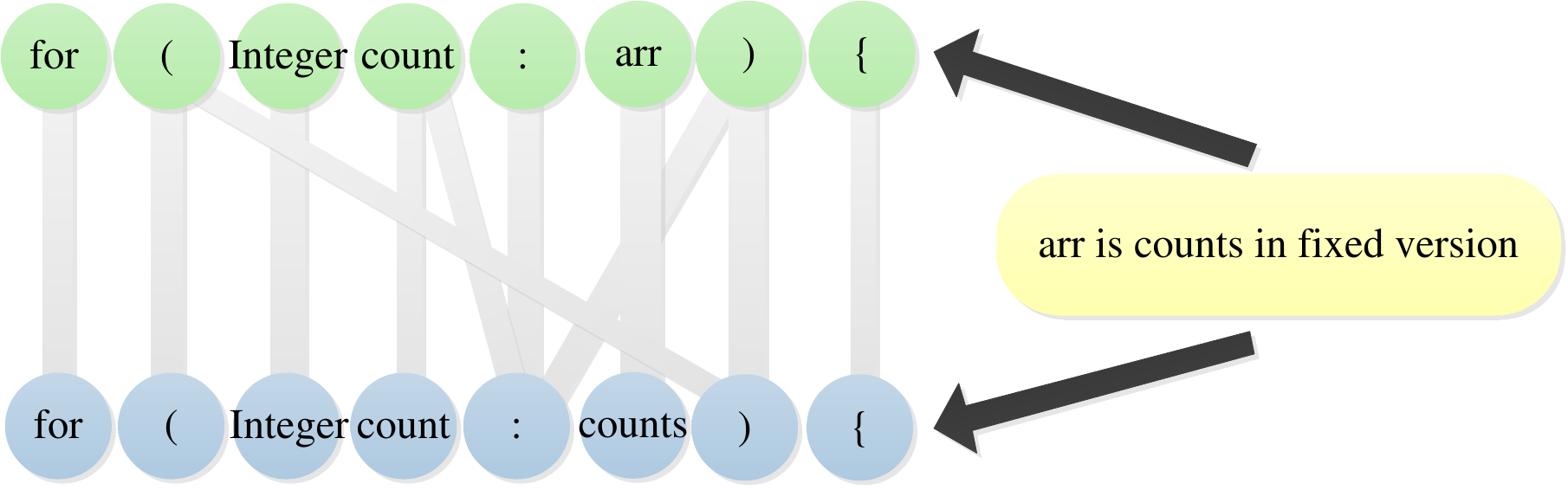}
			\\(c)
		\end{minipage}
		\begin{minipage}[b]{0.49\linewidth}
			\centering
			\includegraphics[width=0.85\linewidth]{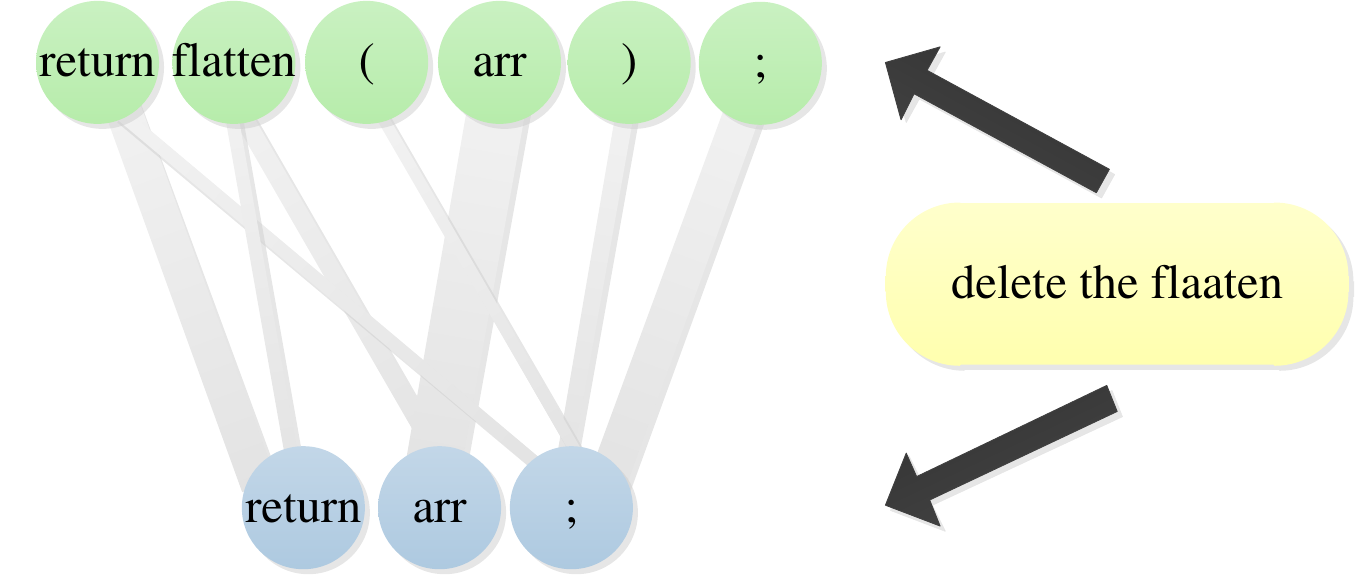}
			\\(d)
		\end{minipage}
		\caption{Dependency estimates and explanation graph generated by our framework. The green circles means the buggy code, the yellow circles means the comments text, the blue circles means debugged code.(a) Explanation graph in Python.(b) Explanation graph in Python.(c) Explanation graph in Java.(d) Explanation graph in Java.}
		\label{fig:exgraphs}
	\end{figure}

	Fig.~\ref{fig:exgraphs} shows the input and output's interpretable relation with explanation graphs. The connected blue circle with more gray lines from green circles is more relevant. The density of gray lines might show whether or not the error code corresponds to the proper code. As shown in Fig.~\ref{fig:exgraphs}, the incorrect part of the buggy code is most connected to the correct part in the repaired code. 
	\vspace{0.4cm}
	
	\begin{figure}[htbp]
		\begin{minipage}[b]{0.49\linewidth}
			\centering
			\includegraphics[width=0.95\linewidth]{fig/explain-other-002-1-1-eps-converted-to.pdf}
			\\(a)
		\end{minipage}
		\begin{minipage}[b]{0.49\linewidth}
			\centering
			\includegraphics[width=0.97\linewidth]{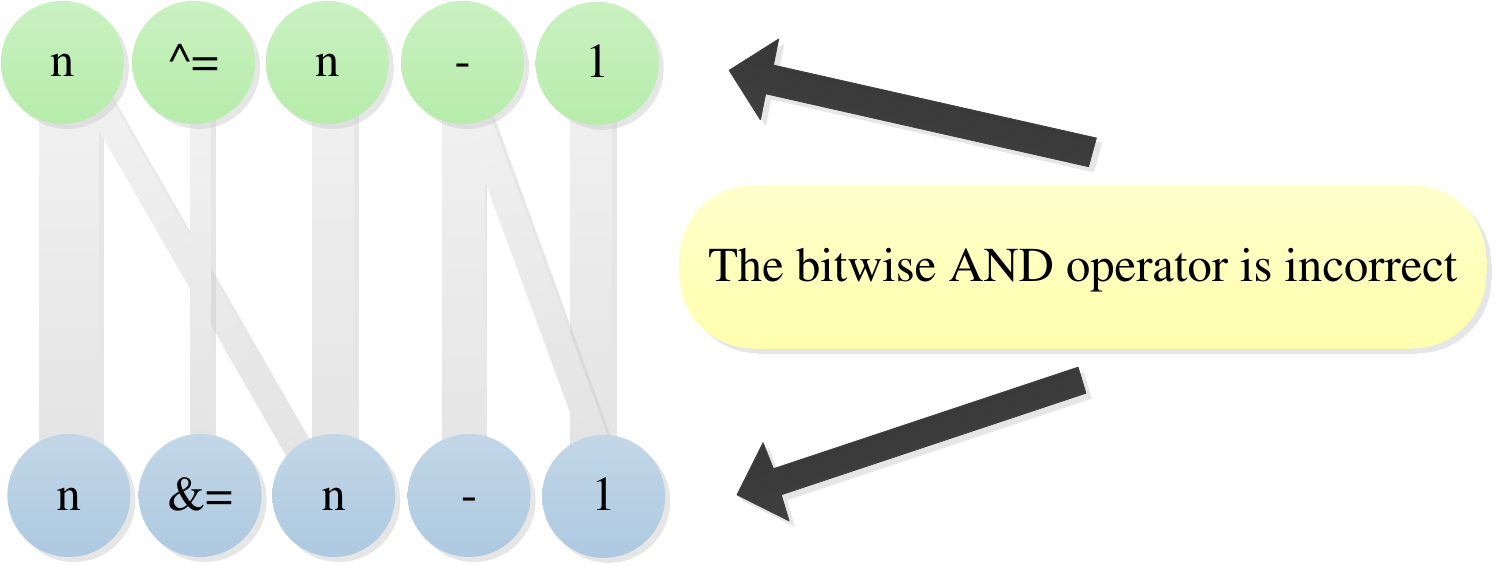}
			\\(b)
		\end{minipage}
		\begin{minipage}[b]{0.49\linewidth}
			\centering
			\includegraphics[width=0.95\linewidth]{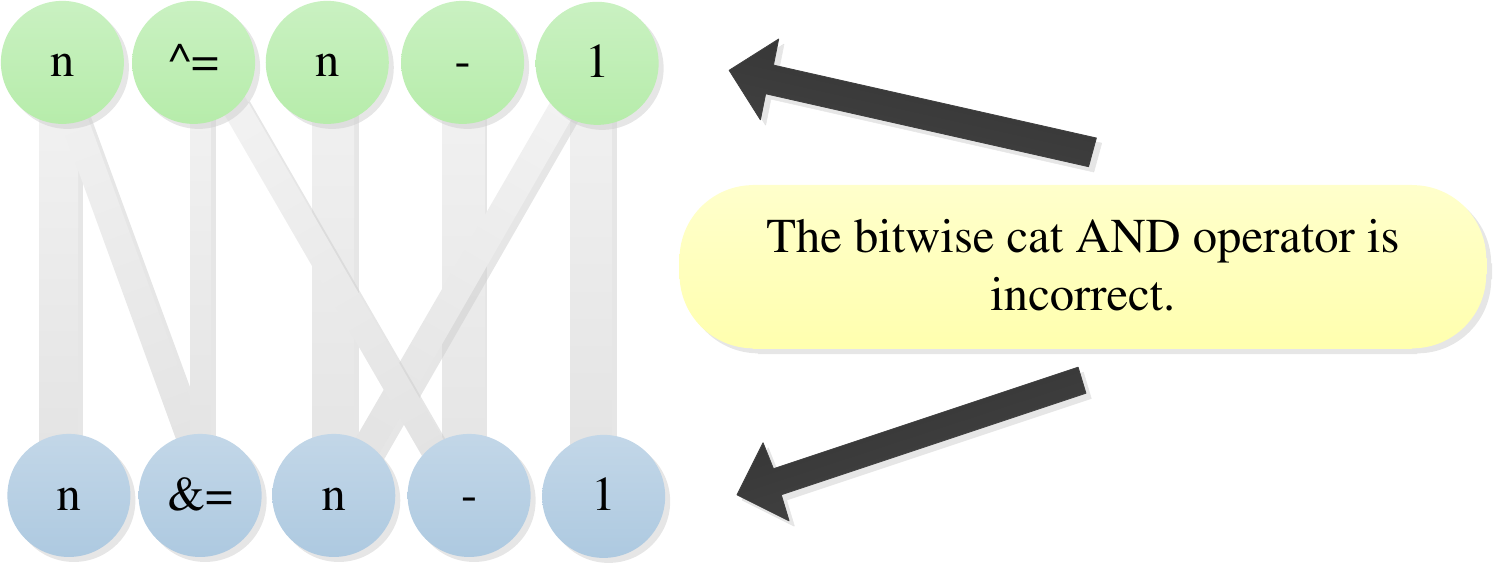}
			\\(c)
		\end{minipage}
		\begin{minipage}[b]{0.49\linewidth}
			\centering
			\includegraphics[width=0.95\linewidth]{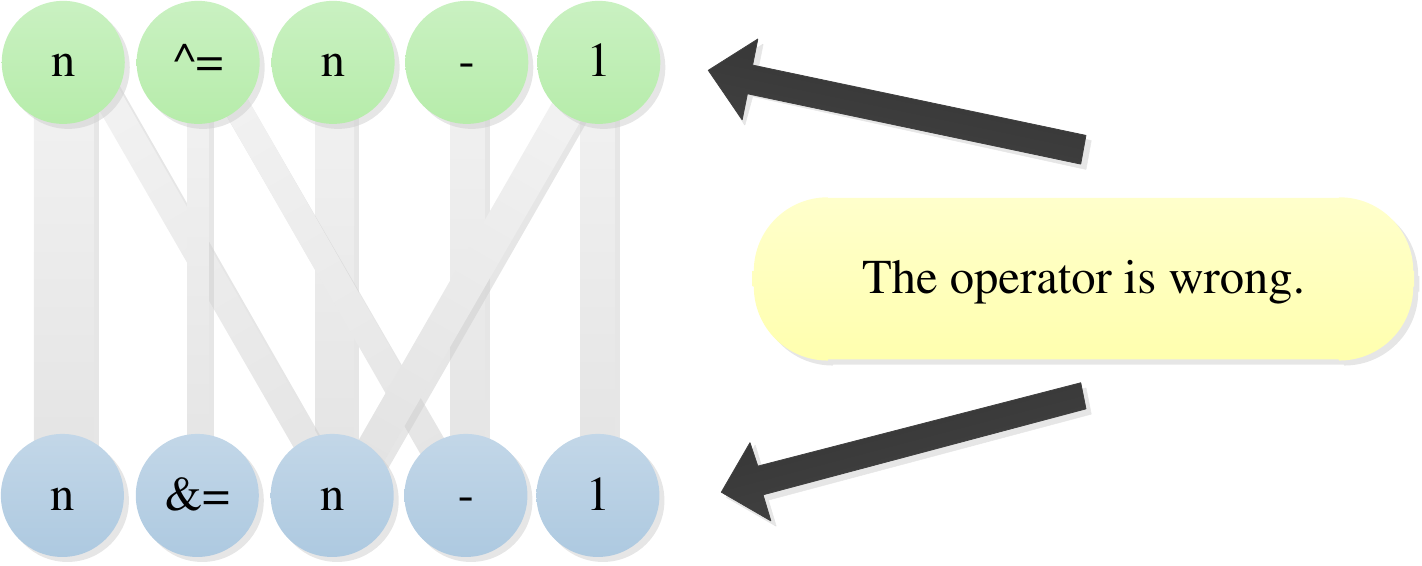}
			\\(d)
		\end{minipage}
		\begin{minipage}[b]{0.49\linewidth}
			\centering
			\includegraphics[width=0.95\linewidth]{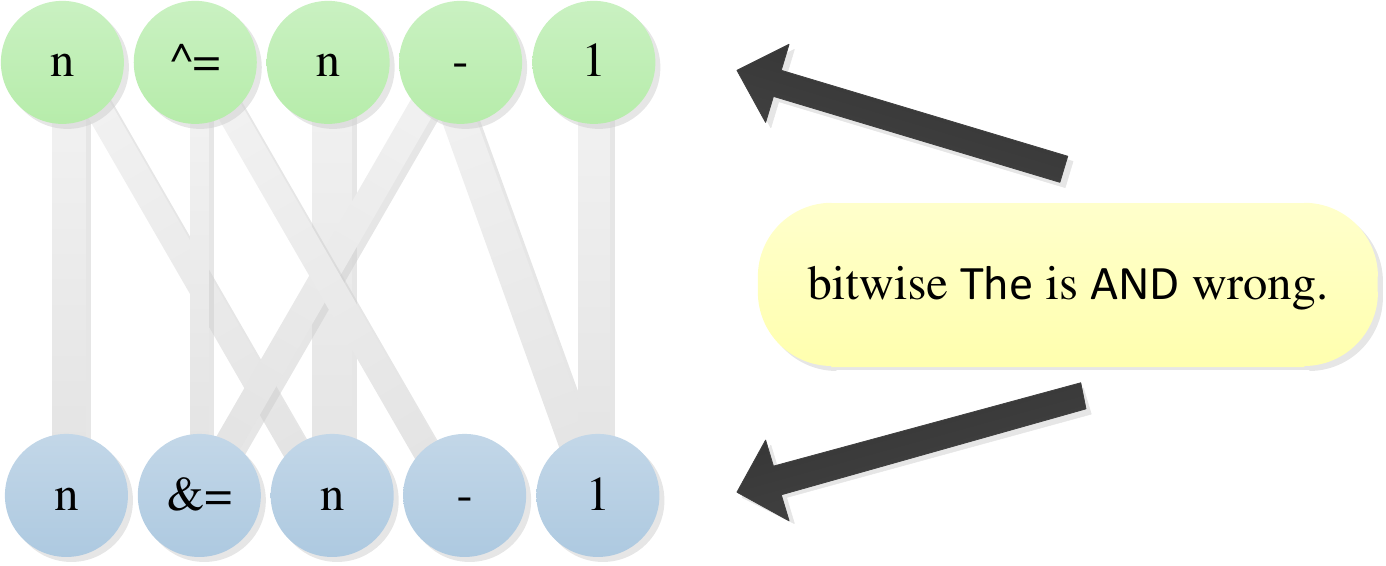}
			\\(e)
		\end{minipage}
		\begin{minipage}[b]{0.49\linewidth}
			\centering
			\includegraphics[width=1.01\linewidth]{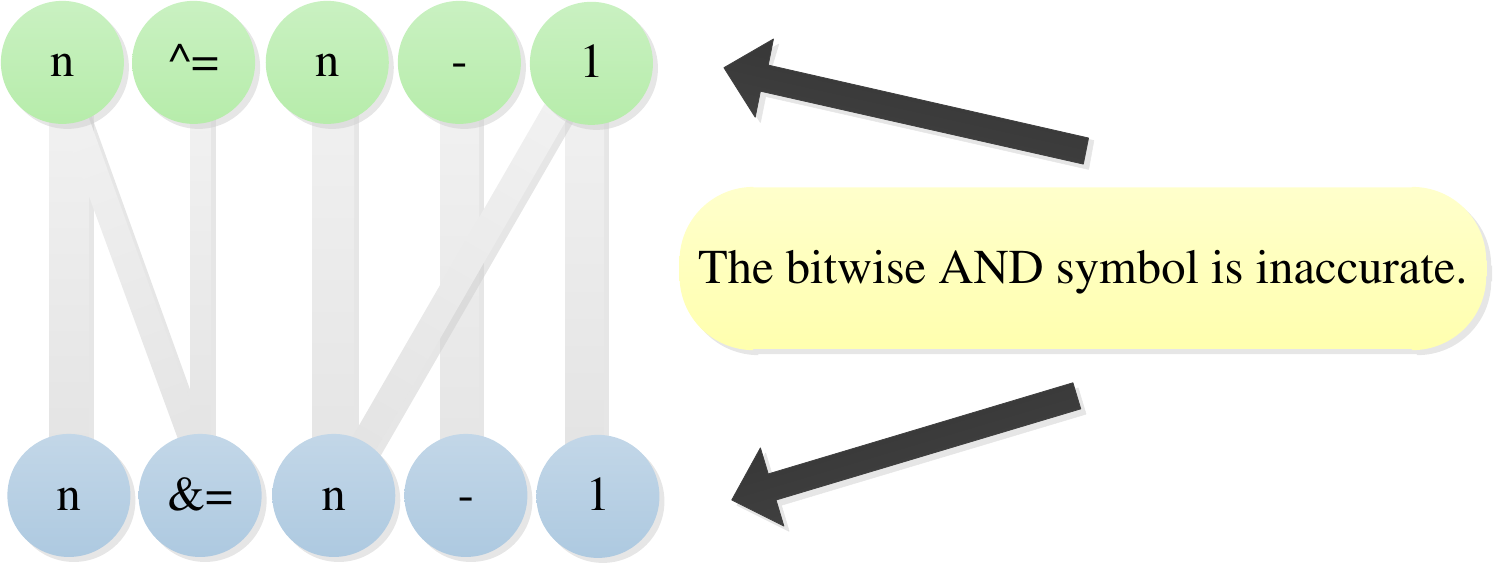}
			\\(f)
		\end{minipage}
		\caption{Dependency estimates and explanation graph generated with data disturbances. The green circles mean the buggy code, the yellow circles mean the comments text, and the blue circles mean debugged code. (a) Explanation graph in Python, without data disturbances. (b) Explanation graph in Python, with SR. (c) Explanation graph in Python, with RI. (d) Explanation graph in Python, with RD. (e) Explanation graph in Python, with RS. (f) Explanation graph in Python, with BT.}
		\label{fig:augexgraphs}
	\end{figure}
	\vspace{0.3cm}
	
	\noindent\textbf{The explanation graph in data perturbation.}
	Fig.~\ref{fig:exgraphs} shows the explanation graph when using data augmentation. In Fig.~\ref{fig:exgraphs} (a), (b), and (c), the buggy code is all connected to the most relevant input. Especially in Fig.~\ref{fig:exgraphs}(b), the incorrect symbol is connected to the correct symbol, where the second blue circle is the right answer. In Fig.~\ref{fig:exgraphs} (d), the debugged code deletes the wrong parts compared to the buggy code. Since the solution is to delete the buggy code, the input is rarely connected to the output.
	
	Fig.~\ref{fig:augexgraphs} (b), (c), (d), (e), and (f) illustrate the circumstance when the data augmentation is implemented for the comment text. It shows that the comment text is still connected to the most relevant part but the irrelevant connections have increased. As in Fig.~\ref{fig:augexgraphs} (c), (d), and (e), the second blue circle gets the connection, but the graph has deleted some unimportant connections to other circles. Data augmentation like RI, RD and RS will cause some irrelevant connections to grow. The SR and BT methods can preserve the correct connections as seen in Fig.~\ref{fig:augexgraphs}. 
	
	Overall, Fig.~\ref{fig:augexgraphs} shows that our framework can demonstrate the interpretability of input and output. The data augmentation strategy and the explanation graph continuously provide improvements to model effects, but appropriate data augmentation approaches will make them more powerful.
	\vspace{0.1cm}
	
	\section{Conclusion}
	Our data augmentation approach combined with a model-agnostic framework for prediction interpretability can produce reasonable, related explanations. We formulate the explanation problem of APR model in causal inference. Our methods can be generalized to a variety of settings in which inputs and outputs can be expressed as sets of features. Also, we used data augmentation sampling for data disturbances. Experiments with various data augmentation methodologies suggest that the APR model can be utilized as a causal inference tool.

	\section*{Acknowledgment}
	This paper is supported by the Key Research and Development Program of Guangdong Province under grant No.2021B0101400003. Corresponding author is Shijing Si from Ping An Technology (Shenzhen) Co., Ltd (sishijing204@pingan.com.cn).

	\small
	\bibliographystyle{IEEEtran}
	\bibliography{IEEEexample}

\begin{thebibliography}{10}
\providecommand{\url}[1]{#1}
\csname url@samestyle\endcsname
\providecommand{\newblock}{\relax}
\providecommand{\bibinfo}[2]{#2}
\providecommand{\BIBentrySTDinterwordspacing}{\spaceskip=0pt\relax}
\providecommand{\BIBentryALTinterwordstretchfactor}{4}
\providecommand{\BIBentryALTinterwordspacing}{\spaceskip=\fontdimen2\font plus
\BIBentryALTinterwordstretchfactor\fontdimen3\font minus
  \fontdimen4\font\relax}
\providecommand{\BIBforeignlanguage}[2]{{%
\expandafter\ifx\csname l@#1\endcsname\relax
\typeout{** WARNING: IEEEtran.bst: No hyphenation pattern has been}%
\typeout{** loaded for the language `#1'. Using the pattern for}%
\typeout{** the default language instead.}%
\else
\language=\csname l@#1\endcsname
\fi
#2}}
\providecommand{\BIBdecl}{\relax}
\BIBdecl

\bibitem{monperrus2018automatic}
M.~Monperrus, ``Automatic software repair: a bibliography,'' \emph{ACM
  Computing Surveys (CSUR)}, vol.~51, no.~1, pp. 1--24, 2018.

\bibitem{sundararaman2021syntactic}
D.~Sundararaman, V.~Subramanian, G.~Wang, S.~Si, D.~Shen, D.~Wang, and
  L.~Carin, ``Syntactic knowledge-infused transformer and bert models,'' in
  \emph{Proceedings of the {CIKM} 2021 Workshops}, 2021.

\bibitem{tufano2018empirical}
M.~Tufano, C.~Watson, G.~Bavota, M.~Di~Penta, M.~White, and D.~Poshyvanyk, ``An
  empirical investigation into learning bug-fixing patches in the wild via
  neural machine translation,'' in \emph{Proceedings of ASE}, 2018, pp.
  832--837.

\bibitem{chen2019sequencer}
Z.~Chen, S.~J. Kommrusch, M.~Tufano, L.-N. Pouchet, D.~Poshyvanyk, and
  M.~Monperrus, ``Sequencer: Sequence-to-sequence learning for end-to-end
  program repair,'' \emph{IEEE Transactions on Software Engineering}, 2019.

\bibitem{li2020dlfix}
Y.~Li, S.~Wang, and T.~N. Nguyen, ``Dlfix: Context-based code transformation
  learning for automated program repair,'' in \emph{Proceedings of ICSE}, 2020,
  pp. 602--614.

\bibitem{lutellier2020coconut}
T.~Lutellier, H.~V. Pham, L.~Pang, Y.~Li, M.~Wei, and L.~Tan, ``Coconut:
  Combining context-aware neural translation models using ensemble for program
  repair,'' in \emph{Proceedings of the 29th ACM SIGSOFT International
  Symposium on Software Testing and Analysis}, 2020, pp. 101--114.

\bibitem{iyer2016summarizing}
S.~Iyer, I.~Konstas, A.~Cheung, and L.~Zettlemoyer, ``Summarizing source code
  using a neural attention model,'' in \emph{Proceedings of ACL}, 2016, pp.
  2073--2083.

\bibitem{ling2016latent}
W.~Ling, P.~Blunsom, E.~Grefenstette, K.~M. Hermann, T.~Ko{\v{c}}isk{\`y},
  F.~Wang, and A.~Senior, ``Latent predictor networks for code generation,'' in
  \emph{Proceedings of ACL}, 2016, pp. 599--609.

\bibitem{yin2017syntactic}
P.~Yin and G.~Neubig, ``A syntactic neural model for general-purpose code
  generation,'' in \emph{Proceedings of ACL}, 2017, pp. 440--450.

\bibitem{nguyen2019automatic}
T.-T. Nguyen, Q.-T. Ta, and W.-N. Chin, ``Automatic program repair using formal
  verification and expression templates,'' in \emph{International Conference on
  Verification, Model Checking, and Abstract Interpretation}.\hskip 1em plus
  0.5em minus 0.4em\relax Springer, 2019, pp. 70--91.

\bibitem{ye2021comprehensive}
H.~Ye, M.~Martinez, T.~Durieux, and M.~Monperrus, ``A comprehensive study of
  automatic program repair on the quixbugs benchmark,'' \emph{Journal of
  Systems and Software}, vol. 171, p. 110825, 2021.

\bibitem{dong2017improving}
Y.~Dong, H.~Su, J.~Zhu, and B.~Zhang, ``Improving interpretability of deep
  neural networks with semantic information,'' in \emph{Proceedings of the IEEE
  CVPR}, 2017, pp. 4306--4314.

\bibitem{falissard2022neural}
L.~Falissard, C.~Morgand, W.~Ghosn, C.~Imbaud, K.~Bounebache, G.~Rey
  \emph{et~al.}, ``Neural translation and automated recognition of icd-10
  medical entities from natural language: Model development and performance
  assessment,'' \emph{JMIR medical informatics}, vol.~10, no.~4, p. e26353,
  2022.

\bibitem{lei2016rationalizing}
T.~Lei, R.~Barzilay, and T.~Jaakkola, ``Rationalizing neural predictions,'' in
  \emph{Proceedings of EMNLP}, 2016, pp. 107--117.

\bibitem{alvarez2017causal}
D.~Alvarez-Melis and T.~Jaakkola, ``A causal framework for explaining the
  predictions of black-box sequence-to-sequence models,'' in \emph{Proceedings
  of EMNLP}, 2017, pp. 412--421.

\bibitem{o2020automatic}
M.~O’Neill and L.~Spector, ``Automatic programming: The open issue?''
  \emph{Genetic Programming and Evolvable Machines}, vol.~21, no.~1, pp.
  251--262, 2020.

\bibitem{liu2020simple}
J.~Liu, Z.~Lin, S.~Padhy, D.~Tran, T.~Bedrax~Weiss, and B.~Lakshminarayanan,
  ``Simple and principled uncertainty estimation with deterministic deep
  learning via distance awareness,'' \emph{NeurIPS}, vol.~33, pp. 7498--7512,
  2020.

\bibitem{hata2018learning}
H.~Hata, E.~Shihab, and G.~Neubig, ``Learning to generate corrective patches
  using neural machine translation,'' \emph{arXiv preprint arXiv:1812.07170},
  2018.

\bibitem{madeiral2019bears}
F.~Madeiral, S.~Urli, M.~Maia, and M.~Monperrus, ``Bears: An extensible java
  bug benchmark for automatic program repair studies,'' in \emph{2019 IEEE 26th
  International Conference on Software Analysis, Evolution and Reengineering
  (SANER)}.\hskip 1em plus 0.5em minus 0.4em\relax IEEE, 2019, pp. 468--478.

\bibitem{tufano2019empirical}
M.~Tufano, C.~Watson, G.~Bavota, M.~D. Penta, M.~White, and D.~Poshyvanyk, ``An
  empirical study on learning bug-fixing patches in the wild via neural machine
  translation,'' \emph{ACM Transactions on Software Engineering and Methodology
  (TOSEM)}, vol.~28, no.~4, pp. 1--29, 2019.

\bibitem{jiang2021cure}
N.~Jiang, T.~Lutellier, and L.~Tan, ``Cure: Code-aware neural machine
  translation for automatic program repair,'' in \emph{Proceedings of
  ICSE}.\hskip 1em plus 0.5em minus 0.4em\relax IEEE, 2021, pp. 1161--1173.

\bibitem{campos2021automatic}
D.~Campos, A.~Restivo, H.~S. Ferreira, and A.~Ramos, ``Automatic program repair
  as semantic suggestions: An empirical study,'' in \emph{Proceedings of
  ICST}.\hskip 1em plus 0.5em minus 0.4em\relax IEEE, 2021, pp. 217--228.

\bibitem{mamatha2022literature}
T.~Mamatha, B.~Reddy, and C.~S. Bindu, ``A literature review on automated code
  repair,'' in \emph{Proceedings of the 2nd International Conference on Recent
  Trends in Machine Learning, IoT, Smart Cities and Applications}.\hskip 1em
  plus 0.5em minus 0.4em\relax Springer, 2022, pp. 249--260.

\bibitem{caruana2015intelligible}
R.~Caruana, Y.~Lou, J.~Gehrke, P.~Koch, M.~Sturm, and N.~Elhadad,
  ``Intelligible models for healthcare: Predicting pneumonia risk and hospital
  30-day readmission,'' in \emph{Proceedings of SIGKDD}, 2015, pp. 1721--1730.

\bibitem{DBLP:conf/cvpr/MahendranV15}
\BIBentryALTinterwordspacing
A.~Mahendran and A.~Vedaldi, ``Understanding deep image representations by
  inverting them,'' in \emph{Proceedings of the IEEE CVPR}.\hskip 1em plus
  0.5em minus 0.4em\relax {IEEE} Computer Society, 2015, pp. 5188--5196.
  [Online]. Available: \url{https://doi.org/10.1109/CVPR.2015.7299155}
\BIBentrySTDinterwordspacing

\bibitem{DBLP:conf/ijcai/ZhaoOK20}
\BIBentryALTinterwordspacing
W.~Zhao, S.~Oyama, and M.~Kurihara, ``Generating natural counterfactual visual
  explanations,'' in \emph{Proceedings of the Twenty-Ninth International Joint
  Conference on Artificial Intelligence, {IJCAI} 2020}, C.~Bessiere, Ed.\hskip
  1em plus 0.5em minus 0.4em\relax ijcai.org, 2020, pp. 5204--5205. [Online].
  Available: \url{https://doi.org/10.24963/ijcai.2020/742}
\BIBentrySTDinterwordspacing

\bibitem{doshi2017roadmap}
F.~Doshi-Velez and B.~Kim, ``A roadmap for a rigorous science of
  interpretability,'' \emph{arXiv preprint arXiv:1702.08608}, vol.~2, p.~1,
  2017.

\bibitem{ribeiro2016should}
M.~T. Ribeiro, S.~Singh, and C.~Guestrin, ``\" why should i trust you?"
  explaining the predictions of any classifier,'' in \emph{Proceedings of
  SIGKDD}, 2016, pp. 1135--1144.

\bibitem{cheng2021fairfil}
P.~Cheng, W.~Hao, S.~Yuan, S.~Si, and L.~Carin, ``Fairfil: Contrastive neural
  debiasing method for pretrained text encoders,'' in \emph{ICLR}, 2021.

\bibitem{DBLP:conf/acl/WuRHW20}
T.~Wu, M.~T. Ribeiro, J.~Heer, and D.~S. Weld, ``Polyjuice: Generating
  counterfactuals for explaining, evaluating, and improving models,'' in
  \emph{Proceedings of {ACL/IJCNLP}}.\hskip 1em plus 0.5em minus 0.4em\relax
  ACL, 2021, pp. 6707--6723.

\bibitem{wei2019eda}
J.~Wei and K.~Zou, ``Eda: Easy data augmentation techniques for boosting
  performance on text classification tasks,'' in \emph{Proceedings of
  EMNLP-IJCNLP}, 2019, pp. 6383--6389.

\bibitem{edunov2018understanding}
S.~Edunov, M.~Ott, M.~Auli, and D.~Grangier, ``Understanding back-translation
  at scale,'' in \emph{Proceedings of EMNLP}, 2018, pp. 489--500.

\bibitem{just2014defects4j}
R.~Just, D.~Jalali, and M.~D. Ernst, ``Defects4j: A database of existing faults
  to enable controlled testing studies for java programs,'' in
  \emph{Proceedings of the 2014 International Symposium on Software Testing and
  Analysis}, 2014, pp. 437--440.

\bibitem{lin2017quixbugs}
D.~Lin, J.~Koppel, A.~Chen, and A.~Solar-Lezama, ``Quixbugs: A multi-lingual
  program repair benchmark set based on the quixey challenge,'' in
  \emph{Proceedings Companion of the 2017 ACM SIGPLAN International Conference
  on Systems, Programming, Languages, and Applications: Software for Humanity},
  2017, pp. 55--56.

\bibitem{le2015manybugs}
C.~Le~Goues, N.~Holtschulte, E.~K. Smith, Y.~Brun, P.~Devanbu, and S.~Forrest,
  et~al., ``The manybugs and introclass benchmarks for automated repair of c
  programs,'' \emph{IEEE Transactions on Software Engineering}, vol.~41,
  no.~12, pp. 1236--1256, 2015.

\bibitem{hanam2016discovering}
Q.~Hanam, F.~S. d.~M. Brito, and A.~Mesbah, ``Discovering bug patterns in
  javascript,'' in \emph{Proceedings of the 2016 24th ACM SIGSOFT international
  symposium on foundations of software engineering}, 2016, pp. 144--156.

\bibitem{li2017hyperband}
L.~Li, K.~Jamieson, G.~DeSalvo, A.~Rostamizadeh, and A.~Talwalkar, ``Hyperband:
  a novel bandit-based approach to hyperparameter optimization,'' \emph{The
  Journal of Machine Learning Research}, vol.~18, no.~1, pp. 6765--6816, 2017.

\bibitem{dhillon2001co}
I.~S. Dhillon, ``Co-clustering documents and words using bipartite spectral
  graph partitioning,'' in \emph{Proceedings of SIGKDD}, 2001, pp. 269--274.

\bibitem{kluger2003spectral}
Y.~Kluger, R.~Basri, J.~T. Chang, and M.~Gerstein, ``Spectral biclustering of
  microarray data: coclustering genes and conditions,'' \emph{Genome research},
  vol.~13, no.~4, pp. 703--716, 2003.

\bibitem{paszke2019pytorch}
A.~Paszke, S.~Gross, F.~Massa, A.~Lerer, J.~Bradbury, and G.~Chanan, et~al.,
  ``Pytorch: An imperative style, high-performance deep learning library.'' in
  \emph{NeurIPS}, 2019.

\bibitem{gehring2017convolutional}
J.~Gehring, M.~Auli, D.~Grangier, D.~Yarats, and Y.~N. Dauphin, ``Convolutional
  sequence to sequence learning,'' in \emph{ICML}.\hskip 1em plus 0.5em minus
  0.4em\relax PMLR, 2017, pp. 1243--1252.

\bibitem{feng2020codebert}
Z.~Feng, D.~Guo, D.~Tang, N.~Duan, X.~Feng, and M.~Gong, et~al., ``Codebert: A
  pre-trained model for programming and natural languages,'' in
  \emph{Proceedings of EMNLP: Findings}, 2020, pp. 1536--1547.

\end{thebibliography}
	
\end{document}